# Artificial Intelligence Edge Applications in 5G Networks

Carlota Villasante Marcos[1][0000-0002-4819-7530]

[1] Ericsson España SA, Madrid, Spain
carlota.villasante@ericsson.com

**Abstract.** In recent years, the 5th Generation of mobile communications has been thoroughly researched to improve the previous 4G capabilities. As opposed to earlier architectures, 5G Networks provide low latency access to services with high reliability. Additionally, they allow exploring new opportunities for applications that need to offload computing load in the network with a real-time response. This paper analyzes the feasibility of a real-time Computer Vision use case model in small devices using a fully deployed 5G Network. The results show an improvement in Latency and Throughput over previous generations, and a high percentage of Availability and Reliability in the analyzed use case.

**Keywords:** 5G, URLLC, Computer Vision, Artificial Intelligence, E2E Latency, OWD, E2E Service Response Time, Availability, Reliability.

## 1 Introduction

Mobile communications have experienced a continuous transformation process, from mobile phone calls and SMS with 2G to video calls and data consumption anywhere with 4G [1]. Nevertheless, the increase in video streaming and high data consuming use cases have caused an impact on the network. In addition, data consumption is rapidly rising, new number of connections appear each day and new technologies as Internet of things (IoT) need better energy efficiency [2]. The Radiocommunication Sector of the International Telecommunication Union (ITU-R) set up a project named IMT-2020 [3], which has established further stricter requirements to provide high speed, high reliability, low latency and energy efficiency mobile services that could not be achieved with previous generations. Within that project, ITU-R defined 5G usage scenarios depending on the requirements addressed on the network: enhanced Mobile BroadBand (eMBB), Ultra-Reliable and Low Latency Communications (URLLC) and massive machine-type communications (mMTC). Every 5G use case or application can be allocated to one or several usage scenarios, depending on the requirements established on the key capabilities of IMT-2020 [4].

The requirements set, and the progress made on the standardization of 5G, offer network operators the possibility to provide new services to a high number of areas such as healthcare, industry, education, tourism, etc [5]. Computer Vision (CV) can be found nowadays in several real examples and use cases (UCs) of those areas, as self-driving car testing, health diagnostics, security and surveillance cameras or even in stores [6]. Not only are Artificial Intelligence (AI) techniques present on new UCs but they also help optimize network performance, enhance customer experience, boost network management by controlling network slicing and solve complex data problems [7]. Most of the solutions with AI rely on heavy machinery, as it requires high computer load for some of the tasks involved. More often, examples of CV can be seen in lighter devices as smartphones or Virtual Reality (VR) glasses, and other personal devices where the workload is done on the device itself, but it still requires quite complex dedicated hardware and software. 5G Networks provide infrastructure and resources, such as edge-computing [8], low latency and high reliability, in order to move the computation load to another location. In this manner, the devices could be lighter, simpler and cheaper, as the intelligence relies on the network and it can become a distributed solution [9]. Offloading the devices' intelligence into the network will also ease to request more computer capacity, fix any encountered problem and update the software more easily, since the software is in a centralized server.

The appearance of so many new UCs has encouraged the launch of several projects and research programs, such as 5G Public-Private Partnership (5GPPP) to explore the possibilities of 5G. Focusing in a European scenario, we can find 5G-EVE "5G European Validation platform for Extensive trials" project [10] whose main purpose is to facilitate a way to validate 5GPPP projects network Key Performance Indicators (KPIs) and services. 5G-EVE evaluates real use cases in a 5G network environment [11] by trying to evaluate which 5GPPP scenario they fit in and the requirements needed to fulfill necessities, showcasing them with a final demo. 5G-EVE gives an end-to-end facility by interconnecting European sites on Greece, Italy, France and Spain. Specifically, in Spain, 5TONIC [12] was designated as the Spanish laboratory site for 5G-EVE, and where the realization of this study takes place.

In this paper, we introduce the analysis of real-time CV with image recognition on small connected devices in 5G networks. For this proposal, a fully deployed 5G Non-Standalone (NSA) network located in the 5G-EVE Spanish site, based on Ericsson's 5G portfolio, is used to give coverage to a device with an incorporated camera which will be continuously streaming video to the CV application located on the network. The results will demonstrate the benefits of the new generation network over the previous mobile generation, regarding latency, reliability and throughput.

## 2   Materials and Methods

### 2.1   Goals

The main goal of this work is to evaluate the feasibility of Computer Vision with image recognition UCs on the new generation of mobile network. As baseline, a camera is used, streaming video through the network to a server where CV image recognition is being applied and a control command is sent back to the device. On the one hand, edge computing, a technique which enables to perform computation offload in the network near the connected devices, is analyzed to check if there is any advantage. On the other hand, we carry out a comparison between 4G and 5G-New

Radio (NR), since the latter ensures low latency and high reliability on the network, enabling real-time response applications.

## 2.2  Environment

The environment for the study is described in **Fig. 1**, composed by a Camera, a 5G NR NSA Customer Premise Equipment (CPE), a 5G NR Radio Access Network (RAN), option3 or NSA NR [13], a Packet Core (PC) with 5G NSA support and a Server where the CV Application (APP), with a trained Neural Network (NN), is running.

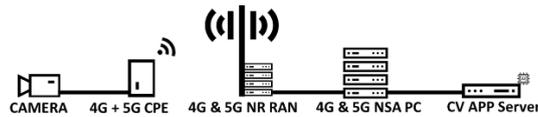

**Fig. 1.** High level environment representation

As the study was done at 5TONIC laboratory, the infrastructure used was provided by the project [14]. As shown in **Fig. 2**, it consists of a raspberry pi with a camera module connected to a 5G CPE as the User Equipment (UE), a RAN (with 4G, 5G NSA and Narrowband-IoT (NB-IoT) access), a Packet Core (with 4G and 5G NSA support) and a Transport layer, which enables to simulate several environments and conditions. Furthermore, the RAN was configured with a bandwidth of 50Mhz in B43(4G)/n78(5G) and a Time-Division Duplexing (TDD) pattern of 7:3 to improve the achievable Uplink throughput.

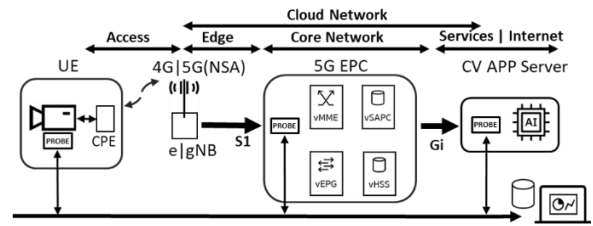

**Fig. 2.** Use case environment representation of 5Tonic equipment.

## 2.3  Measurements

The use case evaluated on this paper is a good example of a URLLC service and the most important KPI is Latency or end-to-end (E2E) Latency, as considered in 5G-EVE [15]. Availability and Reliability, related to the percentage of correctly received packets, must be evaluated. Computer vision use cases usually come alongside with real time response applications where immediate reactions and non-packet-loss are needed. For example, if a pedestrian steps on the way of a vehicle, it needs to react as fast as possible to avoid crashes by slowing down or changing directions. Considering that in these type of UCs the UE needs to send great amount of data through the network, we should also analyze User Data Rate and Peak Data Rate.

E2E Latency as a 5G EVE KPI is measured as the duration between the transmission of a small data packet from a source node and the reception of the corresponding answer at the same point. In this paper, we take several measurement methods of Latency to evaluate the use case, such as Ping in order to evaluate such latency as an emulated control packet, Round Trip Time (RTT), shown in (1), and Smooth RTT (SRTT) in (2), as defined in [16] to measure the E2E Latency on a small TCP packet and on the transmission of a video frame. The latter measurement is considered the most important KPI about latency on this study, as it measures the real E2E Latency of the UC on the network.

$$RTT = Time_{TCP\ ACK} - Time_{TCP\ Packet} \quad (1)$$
$$SRTT = (1 - \alpha) * SRTT + \alpha * R' \quad (2)$$

In addition, One Way Delay (OWD) is also presented. OWD is understood as the time difference between time of transmission and reception. It could be measured in both uplink and downlink, but in this use case the relevant path is uplink, since the video is being sent from the UE to the application layer.

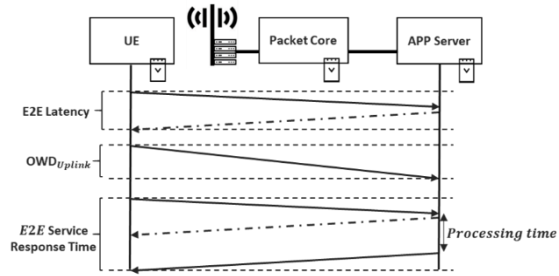

**Fig. 3.** Representation of E2E Latency, OWD and E2E Service Response Time.

Until now, the KPIs correspond to network propagation; however, to calculate the complete E2E Service Response Time and analyze the feasibility of these kind of UCs, the processing time of the video frame on the application must be added, as shown in **Fig. 3**.

### 2.4 Methodology

The User Equipment used on this study is composed of a Raspberry pi 3 model B+ [17] with a camera module [18] connected to a 5G NR CPE. In order to have a live video streaming we used the linux project UV4L [19] with the raspicam driver [20], where several parameters of the video can be set easily, such as, image encoding and resolution, and streaming Server options. In this study, the parameters used were: MJPEG and H264 encoding, 20 frames per second (fps), 640x480 (VGA), 720x576 (D1) and 1280x720 (HD) image resolution [21].

The CV APP Server consists of a dockerized application with a Caffe [22] NN, which receives the images sent by the UE and processes them to perform image recognition to later immediately send a command to the UE. The NN was trained to recognize people and objects that could be found in the lab as shown in **Fig. 4**.

Furthermore, to make the evaluation of the use case and retrieve the measurements and interesting KPIs, the traffic on the network is captured with TCPdump [23], and three specific positions on the environment were selected: the UE, the Core and the APP Server. In order to extract the KPI of E2E latency, the measurement on UE is the only one needed, but to evaluate OWD delay it is necessary to extract the timestamp of the packet in different locations and subtract them, as shown in (3) for a TCP uplink packet.

$$OWD_{(TCP)Uplink} = Time_{(TCP)APP\ Server} - Time_{(TCP)UE} \quad (3)$$

When trying to measure both E2E latency and OWD of a video frame, another dimension appears, since it is not only measured for one packet, but several packets. The video frame can be defined as the group of packets between two packets with a payload value string containing "--Boundary". Despite being TCP, not all the packets have an Acknowledgement (ACK) packet, but instead it is used as a receipt of a group of packets, not necessarily the group corresponding to a video frame. Hence, in this case, the response packet considered to measure E2E Latency is the next ACK packet to appear after a video frame (see **Fig. 5**).

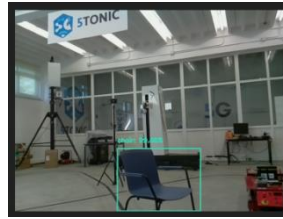

**Fig. 4.** Recognition of a chair shown in the image processed by the CV APP Server at 5Tonic.

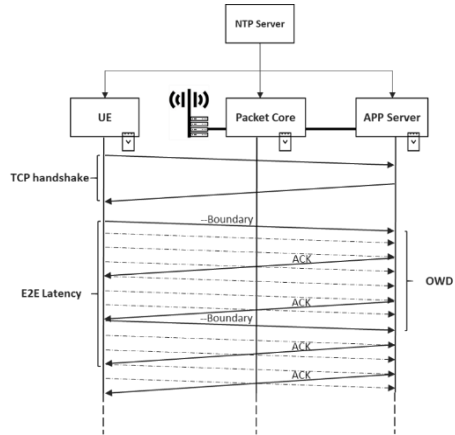

**Fig. 5.** Video frame packet sequence to measure Service E2E Latency and OWD. Representation of synchronization NTP Server-Client mechanism, measurement extraction points on UE, Packet Core and APP Server.

To calculate E2E Service Response Time ($E2E_{SRT}$), the OWD of a video frame in uplink direction $OWDV_U$, the Processing time $\tau$ and time of transmission of the response $OWDR_D$ are needed, represented in (4).

$$E2E_{SRT} = OWDV_U + \tau + OWDR_D \quad (4)$$

To have the minimum possible error and the highest precision, lab synchronization is one of the key components of the study, and we achieved it by using an Network Time Protocol (NTP) Server-Client mechanism where all devices and nodes are set by the same clock.

In addition, we evaluated several scenarios to have a better understanding of the feasibility of this use case on a 5G network when compared to 4G. We used Netem to add additional latency and simulate the location of the Core and APP Server with respect to the RAN. Edge computing, or local range scenario, is a new scenario which is only present in 5G. Below we considered, edge computing as a range of 0 km, regional range scenario of 200km and the national range scenario of 400km. In order to do this analysis and comparison between the results in different scenarios, we added a delay in milliseconds (ms) on the packet core interface, 2ms for a regional scenario and 4ms for national scenario.

The experiments of this work aim to assess the feasibility of a Computer Vision use case in a 5G network. Depending on the application where CV is present, the delay might be crucial, in an automotive scenario that time is directly related with the velocity (v) of a vehicle or the space ($\Delta s$) traveled when braking, as shown in (5).

$$v = \Delta s / \Delta t = \Delta s / E2E_{SRT} \quad (5)$$

## 3    Results

The main measurement on this analysis is E2E Latency. First, in order to set a baseline, we measured Latency with the Ping utility for an ICMP packet, which is considered in this study as a control packet.

Second, we had to consider that the video stream in this use case is sent over Transmission Control Protocol (TCP) and, since it is a connection-oriented protocol, it guarantees that all sent packets will reach destination in order and allows to take trustworthy measurements. We took the first measurements for a video stream with an MJPEG encoder and 640x480 resolution. Both Latency for a control packet and TCP packets are exposed in **Fig. 6**.

Third, as previously explained, it is important to consider the Service E2E latency. This is the case for the transmission of a whole video frame, shown in **Fig. 7**.

Depending on the type of traffic, control packet, TCP packet or whole video frame, the E2E latency values differ, but we can observe that all 5G scenarios have lower values than on both 4G regional and national scenarios. 5G Edge computing scenario is the fastest and it reduces in 23% the latency for the presented UC, 66% for control packets and 44% for TCP packets.

Mentioned in section 2.3, OWD can be measured in both directions: uplink and downlink. However, in this scenario, the important measurement is on uplink as the traffic load and corresponding delay is greater when transmitting video uplink than sending the confirmation of receipt. To make the comparison, we set three different image resolutions (VGA, D1 and HD) with two different encoders (MJPEG and H264) to observe a possible variation on the OWD (see **Fig. 8**).

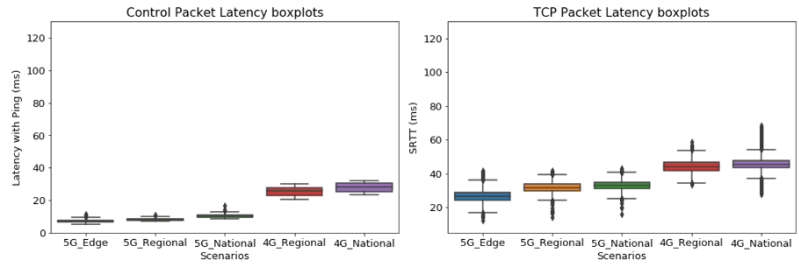

**Fig. 6.** Left: Control Packet Latency boxplot measured with Ping on each scenario.
Right: TCP Packet Latency boxplot measured with SRTT on each scenario.

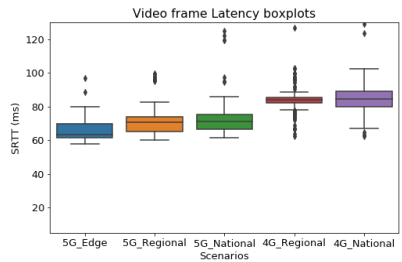

**Fig. 7.** Service Video Frame Latency boxplot measured with SRTT on each scenario.

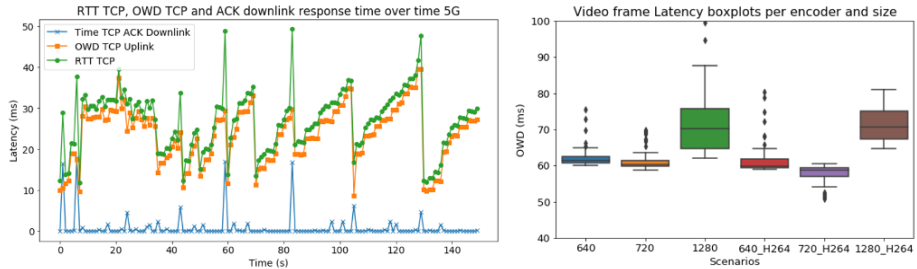

**Fig. 8.** Left: Comparison of RTT, OWD uplink and Ack transmission on a 5G network.
Right: Video Frame OWD boxplot per encoder and size.

As mentioned before, all nodes and devices were synchronized with a common clock in order to achieve the minimum error and higher precision. To calculate the propagation error on this environment, we measured the ntp offset in every node every 10 seconds. If a node is UE, b node is the Core and the dockerized CV APP server is c, we must apply (6) and (7) as defined in [24] to calculate the propagation error. The total error can be approximated as a sum of normal distribution errors.

$$Q = a + b + c \quad (6)$$
$$\sigma_Q^2 = (\sigma_a)^2 + (\sigma_b)^2 + (\sigma_c)^2 \quad (7)$$

For each node, we calculated the standard deviations of the ntp offsets, being 0.387ms on the UE, 0.317ms on the Core and 0.117ms on the Container, which results on a standard propagated error of ±0.821ms.

The CV application consist of different computing actions: transformation of the images received to NN input known as Blob, the NN detection, interpretation of the detections and command creation. During the tests, the processing time was observed to be similar with both MJPEG and H264 encoding and with different image resolutions, a total of 20.3ms. In addition, if we calculate the E2E Service Response Time, as defined in (4), for a MJPEG video stream with VGA resolution and assuming a $OWD_{RD}$ of 5ms, the result is 87ms. In **Fig. 9**, the time distribution of each action in the CV APP and the time distribution of the E2E Response Time can be appreciated.

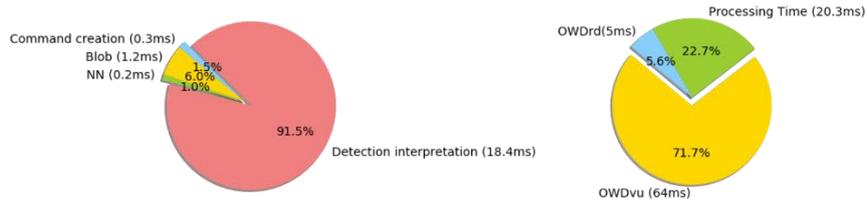

**Fig. 9.** Left: Time distribution of each action performed on the CV application.
Right: Time distribution of E2E Service Response Time.

Furthermore, Availability is defined by the percentage of packets that were successfully delivered through the network. In this study, no packet loss was found which corresponds to a 100% of Availability in 5G.

Regarding Reliability, it is defined as the percentage of successfully delivered packets within the time constraint required by the service. In **Fig. 10**, the Video frame OWD Latency cumulative distribution functions (CDFs) for the VGA resolution with MJPEG encoding can be observed.

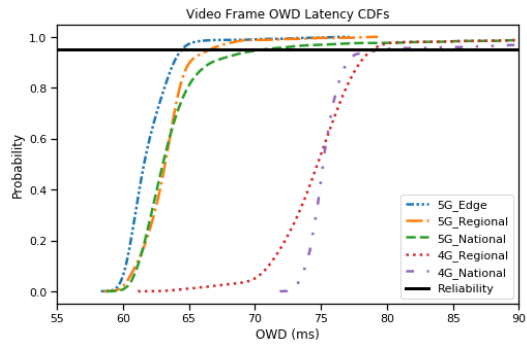

**Fig. 10.** Video frame OWD Latency CDFs for the VGA resolution on each scenario and reliability set to 95%.

Based on (5), the maximum velocity of a vehicle was calculated in both networks and scenarios, considering that it must detect an object and command an order within 1 meter. Assuming E2ESRT as before but with an OWD for upper 95% reliability the results for the calculated velocity are shown in **Table 1**.

Finally, in order to determine the maximum achievable bandwidth, we used iperf3 [25] on both 5G and 4G networks, achieving 54.6Mbits/sec and 32.2Mbits/sec respectively. In **Fig. 11**, the demanded throughput with each type of image resolution is shown. It can be observed, for both D1 and HD resolutions, that the demanded throughput is above the maximum achievable bandwidth in a 4G network, this implies that to maintain a good Quality of Service a downgrade is required.

**Table 1.** Table captions should be placed above the tables.

|  | 5G Edge | 5G Regional | 5G National | 4G Regional | 4G National |
|---|---|---|---|---|---|
| Velocity(km/h) | 40,31 km/h | 39,43 km/h | 37,7 km/h | 35,19 km/h | 34,51 km/h |

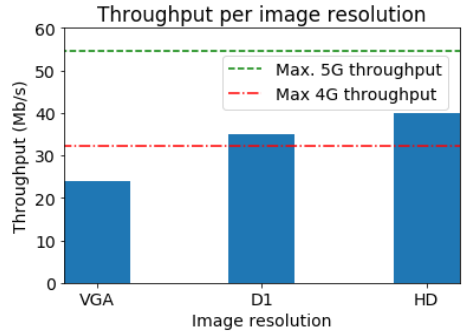

**Fig. 11.** Demanded Throughput per image resolution and Maximum achievable throughput thresholds.

## 4      Conclusions

This paper presents the improvements of 5G NSA for a Computer Vision model UC with image resolution. As appreciated on the results, all explored scenarios on 5G have lower values on the calculated E2E latency for each type of traffic, being Edge computing the fastest one.

In this UC, the predominant direction of traffic is found to be the uplink traffic and a slight difference is appreciated on the OWD results when the resolution of the video is changed. Despite the processing time on the calculation of the E2E Service Response time, it is observed that 5G Edge scenarios allow a faster velocity on a vehicle, in order to detect an object within 1 meter and being able to react with a reliability higher than 95% in the network.

In addition, the maximum throughput achievable in a 4G network is below the observed demanded throughput for higher resolutions. This implies that a 5G network is needed to obtain a video stream with high resolutions and, at least, 20fps.

**Acknowledgements.** This Master Thesis was supported by Ericsson and Universidad Carlos III de Madrid. I would like to thank Marc Mollà Roselló for his assistance as the Thesis supervisor, and the colleagues from 5G-EVE project who provided insight and expertise that assisted the research. Also, the friends and family who provided comments that greatly improved the manuscript.